\documentclass[aos,MSNbibl,seceqn,citesort,dvips]{arximspdf}
\usepackage{accents}

\doi{10.1214/11-AOS933}
\volume{39}
\issue{6}
\pubyear{2011}
\firstpage{3211}
\lastpage{3233}

\makeatletter

\renewcommand{\mathring}[1]{\accentset{\circ}{#1}}

\newtheorem{theorem}{Theorem}

\newproclaim{algorithm}{Algorithm}

\newtheorem{lemma}{Lemma}

\newproclaim{condition}{Condition}

\newcommand{\C}{{\mathbb C}}

\makeatother

\begin{document}
\begin{frontmatter}

\title{An asymptotic error bound for testing multiple quantum hypotheses}
\runtitle{Testing multiple quantum hypotheses}

\begin{aug}
\author[A]{\fnms{Michael} \snm{Nussbaum}\thanksref{t1}\ead[label=e1]{nussbaum@math.cornell.edu}}
\and
\author[B]{\fnms{Arleta} \snm{Szko\l a}\corref{}\thanksref{t2}\ead[label=e2]{szkola@mis.mpg.de}\ead[label=u1,url]{http://personal-homepages.mis.mpg.de/szkola/}}
\runauthor{M. Nussbaum and A. Szko\l a}
\affiliation{Cornell University and Max Planck Institute}
\address[A]{Department of Mathematics\\
Malott Hall\\
Cornell University\\
Ithaca, New York 14853\\
USA\\
\printead{e1}}
\address[B]{Max Planck Institute\\
\quad for Mathematics in the Sciences\\
Inselstrasse 22\\
04103 Leipzig\\
Germany\\
\printead{e2}\\
\printead{u1}} %adresu isvedimo komanda gale!
\end{aug}

\thankstext{t1}{Supported by NSF Grants DMS-08-05632 and DMS-11-06460.}

\thankstext{t2}{Supported by German Research Foundation (DFG)
via the project ``Quantum Statistics: Decision
Problems and Entropic Functionals on State Spaces.''}

% HISTORY:
\received{\smonth{7} \syear{2011}}

% ABSTRACT
%
\begin{abstract}
We consider the problem of detecting the true quantum state among $r$
possible ones, based of measurements performed on $n$ copies of a
finite-dimensional quantum system. A special case is the problem of
discriminating between $r$ probability measures on a finite sample
space, using $n$ i.i.d. observations. In this classical setting, it is
known that the averaged error probability decreases exponentially with
exponent given by the worst case binary Chernoff bound between any
possible pair of the $r$ probability measures. Define analogously the
multiple quantum Chernoff bound, considering all possible pairs of
states. Recently, it has been shown that this asymptotic error bound is
attainable in the case of $r$ pure states, and that it is unimprovable
in general. Here we extend the attainability result to a larger class
of $r$-tuples of states which are possibly mixed, but pairwise
linearly independent. We also construct a quantum detector which
universally attains the multiple quantum Chernoff bound up to a
factor $1/3$.
\end{abstract}

% KEYWORDS
%

\end{frontmatter}

%s1 #&#
\section{Introduction}

Consider a finite set $\Sigma=\{ P_{1},\ldots,P_{r}\} $ of
probability distributions on a sample space $\Omega$, and the problem of
discriminating between them on the basis of observed i.i.d. data. It is well
known that for the maximum likelihood decision rule, the error probability
(Bayesian for uniform prior) decreases exponentially, with a rate given
by the
worst case among the possible pairwise hypothesis testing problems.
Indeed if
$\xi_{\mathrm{CB}}(P_{i},P_{j})$ represents the rate of exponential decay of
the error
probability for deciding between $P_{i}$ and $P_{j}$, given by the classical
Chernoff bound
\[
\xi_{\mathrm{CB}}(P_{i},P_{j})=-\log\inf_{0\leq s\leq1}\int(
dP_{i})
^{1-s}( dP_{j}) ^{s},
\]
then the \textit{multiple Chernoff bound} pertaining to the set
$\Sigma$ has
been defined~as
%
%e1.1 #&#
\begin{equation} \label{class-CB-def}
\xi_{\mathrm{CB}}(\Sigma):=\min\{ \xi_{\mathrm{CB}}(P_{i},P_{j})\dvtx
P_{i},P_{j}
\in\Sigma,P_{i}\neq P_{j}\}
\end{equation}
(Salikhov \cite{Salikh-73,Salikh-97,Salikh-2002}).
If $\pi_{n}$ is the
maximum likelihood rule for sample size $n$, with values in $\{
1,\ldots,r\} $, then, under a uniform prior on $\Sigma$
%
%e1.2 #&#
\begin{equation}\label{class-multip-Chern}
-\frac{1}{n}\log\Pr( \pi_{n}\neq i) \rightarrow\xi_{\mathrm{CB}}
(\Sigma)\qquad\mbox{as }n\rightarrow\infty
\end{equation}
and since $\pi_{n}$ is also Bayesian here, the quantity
$\xi_{\mathrm{CB}}(\Sigma)$ is the best possible asymptotic error exponent for
any decision rule under a uniform prior.

\textit{On terminology}. When large deviation type limits are written
in logarithmic form as in (\ref{class-multip-Chern}), then the right-hand
side
$\xi_{\mathrm{CB}}(\Sigma)$ is referred to as the \textit{rate of exponential
decay} or, in information theory, as the \textit{asymptotic error
exponent}, to be maximized by decision rules. Throughout the paper, we
adhere to this formulation as a convenient equivalent to minimizing
asymptotic error.

We consider here the analogous problem in a quantum statistical
setting, where $\Sigma=\{ \rho_{1},\ldots,\rho_{r}\} $
is a
set of density operators on the finite-dimensional complex Hilbert
space $\C^{d}$. Recall that by definition a density operator
$\rho$, describing the state of a physical system, is a complex,
self-adjoint, positive semidefinite matrix satisfying the
normalization condition $\operatorname{tr}[\rho]=1$. If all
operators $\rho_{i}\in$ $\Sigma$ commute, then the corresponding
matrix representations are jointly diagonizable, and the problem
becomes one of discriminating between the associated finite
probability distributions appearing on the matrix diagonal.

The starting point for our investigation is the recent extension of the
Chernoff binary testing bound to the quantum setting
\cite{NSz,Audenaert,ANSV}. In full analogy to the classical
case, the
quantum Chernoff bound specifies the asymptotic error in the decision problem
between $\rho_{i}$ and $\rho_{j}$, based on a~rule using the outcomes of
measurements performed on $n$ copies of the basic quantum system.

The case of multiple hypotheses ($r>2$) represented by quantum states has
received some interest in the literature over the past three decades; cf.
\cite{Holevo-JMA-73,Holevo-74,Belavkin,buch-Helstrom,Yuen,parth-extr,parth-ml}
and overviews in \cite{Bergou-Herzog,Chefles,buch-Hayashi}. While in
the binary
case ($r=2$) the optimal quantum test is described explicitly by the
Holevo--Helstrom projections, in the case $r>2$ only an implicit
description in terms of an extremal problem is available (Holevo
\cite{Holevo-JMA-73}, Yuen, Kennedy, Lax \cite{Yuen}). Parthasarathy
\cite{parth-ml} has dubbed the quantum Bayes rule ``quantum maximum
likelihood,'' in view of the fact that in the classical case, for a
finite number of hypotheses, the Bayes rule for uniform prior is
indeed maximum likelihood.

Numerous new contributions to multiple quantum hypothesis testing
appeared in the very recent past, for example,
\cite{Assalini,Hwang,Barett,Tyson,Tyson-10,Gen,KRS,Qiu-10,Montan}. The
main focus has been on characterizing the Bayes rule of
\cite{Holevo-JMA-73,Yuen} and finding approximations to it. We focus
here on the asymptotics of the error probability based on measurements
performed on $n$ of copies of the basic quantum system. The true state
is thus described by the $n$th tensor power~$\rho_{i}^{\otimes n}$ of
one of the original density operators $\rho_{i}\in\Sigma$.
Parthasarathy~\cite{parth-ml} established consistency of the Bayes rule
and also an exponential rate of decay of the error probability, without
specifying the error exponent. The first step toward finding the
\textit{optimal} asymptotic error, for which a similar structure as in
the classical case (\ref{class-multip-Chern}) was conjectured, was made
in~\cite{NSz2}. It was shown that if all $\rho_{i}$ are pure states
[$\operatorname{rank}( \rho_{i}) =1$], then the optimal asymptotic
error is given by $\xi_{\mathrm{QCB}}(\Sigma)$, defined as the worst
case error for quantum discrimination between any pair of distinct
states involved. Thus, the situation is indeed analogous to the
classical case (\ref{class-multip-Chern}), and the quantity
$\xi_{\mathrm{QCB}}(\Sigma)$ describing the asymptotics of the error
probability should be termed the \textit{multiple quantum Chernoff
bound}.

The fact that $\xi_{\mathrm{QCB}}(\Sigma)$ is valid as a lower error bound is
relatively straightforward to prove; for a precise statement of the
result from~\cite{NSz2}; cf. Theorem~\ref{thmNSz2}. Attainability for
pure states has been shown in~\cite{NSz2} by constructing a~measurement
based on a Gram--Schmidt orthonormalization of the~$r$ unit
vectors representing the $\rho_{i}\in\Sigma$. It should be mentioned
that earlier Holevo~\cite{Holevo} showed such a measurement to be an
approximation to the Bayes rule. In~\cite{exp-decay}, it was shown that
without any restriction on the nature of the states, an asymptotic
error $\xi_{\mathrm{QCB}}(\Sigma)$ is achievable up to a factor which is
between $2/r(r-1)$ and $1$, for $r$ being the number of hypotheses.

In the present paper, we develop a new decision rule generalizing two
known asymptotically optimal ones, in the following sense: if all
states commute, the method reduces to classical maximum likelihood (as
does the Bayes rule of \mbox{\cite{Holevo-JMA-73,Yuen}}). If all
states are pure, then it coincides with the orthonormalization
algorithm of \cite{NSz2}. We establish that this rule attains
asymptotic error $\xi_{\mathrm{QCB}}(\Sigma)$ for a class of $r$-tuples of
states which fulfill Condition \ref{condLI} below. The condition allows for
mixed states but excludes faithful ones (full rank density
matrices). We then show that a modified version of our rule is near
optimal, in the sense that it attains at least
$\frac{1}{3}\xi_{\mathrm{QCB}}(\Sigma)$ universally.

The outline of our paper is as follows. In
Section \ref{secpreliminaries}, we introduce notation, specify the
mathematical framework, and state precisely our main results in
Theorems~\ref{thmmultiple-Chernoff} and \ref{thmexp-decay}. Some
further discussion of the quantum Bayes rule, of results in statistics
resembling the multiple Chernoff bound and other topics follows at the
end of that section. In Section \ref{secalgorithm}, our new quantum
decision rule is developed, along with Lemma
\ref{lem-general-algorithm} providing a basic error bound.
Section \ref{secachievability} treats the case of pairwise linearly
independent states [Condition \ref{condLI} and Theorem \ref{thmmultiple-Chernoff}].
Section \ref{seccommute} shows how our decision rule reduces to
maximum likelihood
in the commuting case, such that Lemma \ref{lem-general-algorithm} reproduces
the multiple Chernoff bound of \cite{Salikh-73,Salikh-97}.
Section \ref{secnear-optimal} concerns the general attainability of
the near
optimal error bound (Theorem \ref{thmexp-decay}).

%s2 #&#
\section{Notation and preliminaries}
\label{secpreliminaries}

We will describe here the formalism for the simplest possible
nonclassical setup of discrimination between several quantum
hypotheses. A \textit{density matrix} $\rho$ is a complex,
self-adjoint, positive, $d\times d$ matrix satisfying the
normalization condition $\operatorname{tr}[\rho]=1$, where
$\operatorname{tr}[\cdot] $ is the trace operation. Here,
``positive'' means nonnegative definite. We identify a $d\times d$
density matrix with a \textit{quantum state} on $\mathbb{C}^{d}$; we
also use ``matrix'' and ``operator'' interchangeably. The $r$ hypotheses
are described by states $H_{i}\dvtx\rho=\rho_{i}$,
$i=1,\ldots,r$. Physically discriminating between these states
corresponds to performing a measurement on the quantum system.
Mathematically a quantum decision rule with $r$ possible outcomes is a
set of complex self-adjoint positive matrices $d\times d$ matrices
$E=\{ E_{1},\ldots,E_{r}\} $ satisfying
$\sum_{i=1}^{r}E_{i}=\mathbf{1}$ where $\mathbf{1}$ is the unit
matrix. The $r$-tuple $E$ is often called a~POVM (positive operator
valued measure); we will refer to it as a~\textit{quantum multiple
test} or a \textit{quantum detector.} In the special case where all
$E_{i}$ are projections, the $r$-tuple $E$ is called a PVM (projection
valued measure) or von Neumann measurement. The
\textit{individual success probability}, that is, the probability to
accept hypothesis $H_{i}$ when $\rho_{i}$ is the true state, is given
by\looseness=-1
\[
\mathrm{Succ}_{i}(E):=\mathrm{tr}[\rho_{i}E_{i}].
\]\looseness=0
The corresponding individual error probability, that is, the
probability of
rejecting the true state $\rho_{i}$ according to the decision rule, is
\begin{eqnarray*}
\mathrm{Err}_{i}(E) & = & 1-\mathrm{Succ}_{i}(E)
=\mathrm{tr}[\rho_{i}(\mathbf{1}-E_{i})]\\
&=&\sum_{j=1,j\not=i}
^{r} \mathrm{tr}[\rho_{i}E_{j}].
\end{eqnarray*}
The total (averaged) error probability is then
\[
\operatorname{Err}(E):=\frac{1}{r}\sum_{i=1}^{r}\mathrm{Err}
_{i}(E)=\frac{1}{r}\sum_{i=1}^{r} \mathrm{tr}[\rho_{i}(\mathbf{1}
-E_{i})].
\]
The above describes the basic setup where the finite dimension $d$ is
arbitrary and the hypotheses are equiprobable. We consider the quantum
analog of having $n$ i.i.d. observations. For this, the $r$ hypotheses
are assumed to be~$\rho_{i}^{\otimes n}$, $i=1,\ldots,r$, where
$\rho^{\otimes n}$ is the $n$-fold tensor product of $\rho$ with
itself (a~$d^{n}\times d^{n}$ matrix). The detectors $E=\{
E_{1},\ldots,E_{r}\} $ now operate on the states~$\rho_{i}^{\otimes n} $,
that is, the dimension of the components $E_{i}$
is $d^{n}\times d^{n}$, but $E_{i}$ need not have tensor product
structure. The corresponding total error probability of a detector $E$
is now
\[
\mathrm{Err}_{n}(E)=1-\sum_{i=1}^{r}\frac{1}{r}
\operatorname{tr}[
\rho_{i}^{\otimes n}E_{i}].
\]
For the case of two hypotheses $r=2$, the Bayes test for each
$n\in\mathbb{N}$ is known to be the \textit{Holevo--Helstrom
hypothesis\vadjust{\goodbreak}
test}. It is given by the detector $E_{(n)}^{\ast}=\{
\mathbf{1}-\Pi_{n}^{\ast},\Pi_{n}^{\ast}\} $ where
\[
\Pi_{n}^{\ast}=\operatorname{supp}(\rho_{2}^{\otimes n}-\rho
_{1}^{\otimes
n})_{+},
\]
where $\operatorname{supp} a$ is the projection onto the space spanned by
the columns of~$a$ and $a_{+}$ denotes the positive part of a
self-adjoint operator $a$. Thus, if $a=\sum_{i}\lambda_{i}S_{i}$ is the
spectral decomposition using projections $S_{i}$, then
$a_{+}:=\sum_{\lambda_{i} >0}\lambda_{i}S_{i}$ and $\operatorname{supp}
a_{+}=\sum_{\lambda_{i}>0}S_{i}$. The Bayes test is unique up to a
possible reassignment of the projections $S_{i}$ corresponding to zero
eigenvalues of $a=\rho_{2}^{\otimes n}-\rho_{1}^{\otimes n}$. For
$r>2$, the Bayes detector has been described in
\cite{Holevo-JMA-73,Yuen}. Explicit expressions for its $r$ components are not known
in general; for the convenience of the reader, we present the
available implicit description below at the end of this section.

If for a sequence of detectors $E_{(n)}$ the limit $\lim_{n\rightarrow
\infty }-\frac{1}{n}\log\mathrm{Err}_{n}( E_{(n)}) $ exists, we refer
to it as the (\textit{asymptotic}) \textit{error exponent}. For two
density matrices $\rho_{1}$ and~$\rho_{2}$, the \textit{quantum
Chernoff bound} is defined by
%
%e2.1 #&#
\begin{equation} \label{defq-Chernoff-dist}
\xi_{\mathrm{QCB}}(\rho_{1},\rho_{2}):=-\log\inf_{0\leq s\leq1}
\operatorname{tr}
[ \rho_{1}^{1-s}\rho_{2}^{s}].
\end{equation}
The basic properties of $\xi_{\mathrm{QCB}}(\rho_{1},\rho_{2})$ have been
discussed in \cite{ANSV}. Some distan\-ce-like properties have been
noted by Calsamiglia et al. \cite{calsa-1}. For the binary
discrimination problem, it is known that the Holevo--Helstrom (Bayes)
detector~$E_{(n)}^{\ast}$ satisfies
\[
\lim_{n\rightarrow\infty}-\frac{1}{n}\log\mathrm{Err}_{n}\bigl( E_{(n)}
^{\ast}\bigr) =\xi_{\mathrm{QCB}}(\rho_{1},\rho_{2}),
\]
thus specifying $\xi_{\mathrm{QCB}}(\rho_{1},\rho_{2})$ as the optimal error
exponent (cf. \cite{NSz,Audenaert,ANSV}), and
providing the quantum analog of the classical Chernoff bound,
that is,~(\ref{class-multip-Chern}) for $r=2$.

For a set $\Sigma=\{ \rho_{1},\ldots,\rho_{r}\} $ of
density operators on $\mathbb{C}^{d}$, where $r\geq2$, we have
introduced in \cite{NSz2} the \textit{multiple quantum Chernoff bound}
$\xi_{\mathrm{QCB}}(\Sigma) $
%
%e2.2 #&#
\begin{equation}\label{defq-multiple-Chernoff-dist}
\xi_{\mathrm{QCB}}(\Sigma):=\min\{\xi_{\mathrm{QCB}}(\rho_{i},\rho_{j})\dvtx 1\leq
i<j\leq r\}.
\end{equation}
If all the states are jointy diagonizable (commuting), then
(\ref{defq-multiple-Chernoff-dist}) reduces to the classical multiple
Chernoff bound (\ref{class-CB-def}), as it was defined in
\cite{Salikh-73,Salikh-2002} for hypotheses represented by
probability distributions. Taking the minimum over different pairs of
hypotheses corresponds to the worst case in any of the associated
binary hypothesis testing problems. The following well-known result
shows that $\xi_{\mathrm{QCB}}(\Sigma)$ as a rate exponent cannot be exceeded
(cf. \cite{NSz2}, Theorem 1).
%
%th1 #&#
\begin{theorem}
\label{thmNSz2} Let $\Sigma=\{ \rho_{1},\ldots,\rho
_{r}\} $ be a
finite set of hypothetic states on~$\mathbb{C}^{d}$. Then for any
sequence $\{E_{(n)}\}_{n\in\mathbb{N}}$ of quantum detectors relative
to~$\Sigma^{\otimes n}$, respectively, one has
%
%e2.3 #&#
\begin{equation} \label{ineqNSz2}
\limsup _{n\rightarrow\infty}-\frac{1}{n}\log\mathrm
{Err}_{n}\bigl(
E_{(n)}\bigr) \leq\xi_{\mathrm{QCB}}(\Sigma).
\end{equation}
\end{theorem}

The above theorem has been extended in \cite{exp-decay} to the case of quantum
hypotheses which correspond to identically distributed but not necessary
independent observations. The corresponding upper bound in (\ref{ineqNSz2})
is then replaced by a mean generalized Chernoff distance, as introduced in
\cite{hiai-mosonyi-ogawa} for a stationary observation scheme in
binary case. In \cite{exp-decay}, it was also shown, again in a wider
model corresponding to a
class of correlated observations, that quantum detectors with an exponential
decay of $\mathrm{Err}_{n}( E_{(n)}) $ can be
constructed, with
error exponent $\phi\xi_{\mathrm{QCB}}(\Sigma)$ where $2/{r(r-1)\leq\phi
\leq1}$. The
method used in \cite{exp-decay} yields a factor $\phi$ which may be
close to
one for special ensembles of states, but the guaranteed factor $2/r(r-1)$
decreases with the number of hypotheses.

The following two theorems represent our main results. The support~$\operatorname{supp} (\rho)$ of a state $\rho$ is the subspace of
$\mathbb{C}^{d}$ spanned by its columns. Consider:
\renewcommand{\thecondition}{(LI)}
\begin{condition}\label{condLI}
$\operatorname{supp}(\rho_{i})\cap\operatorname{supp}(\rho
_{j})=\{0\}$ for all $i\neq j$.
\end{condition}

The condition is equivalent to requiring that $\rho_{i}$ and $\rho_{j}$
are linearly independent, in the sense that for any two bases of
$\operatorname{supp} (\rho_{i})$ and $\operatorname{supp} (\rho_{j})$,
the union set of vectors is linearly independent. This is obviously
fulfilled for a set $\Sigma$ of $r$ distinct pure states, but the
condition allows for mixed states if $d > 2$. Indeed,
Condition~\ref{condLI} restricts the dimension of the supports
$\operatorname{supp}(\rho_i)$ according to the inequality
$\operatorname{supp}(\rho_i) + \operatorname{supp}(\rho_j)\leq d$ that
is valid for all $i\not=j$. However, as long as none of the density
matrices is of full rank, that is, rank equal to $d$, no constraints on
the number $r$ of distinct hypothetic states are imposed by Condition
\ref{condLI}.
%
%th2 #&#
\begin{theorem}
\label{thmmultiple-Chernoff} Let $\Sigma$ be a finite set of states on
$\mathbb{C}^{d}$ fulfilling Condition~\ref{condLI}. Then there exists a
sequence $\{E_{(n)}\}_{n\in\mathbb{N}}$ of quantum detectors relative
to $\Sigma^{\otimes n}$, respectively, such that
\[
\lim_{n\rightarrow\infty}-\frac{1}{n}\log\mathrm{Err}_{n}\bigl(
E_{(n)}\bigr) =\xi_{\mathrm{QCB}}(\Sigma).
\]
\end{theorem}

Due to the following theorem in the i.i.d. situation---as considered in the
present paper- an error exponent of $\frac{1}{3}\xi_{\mathrm{QCB}}(\Sigma)$
can always
be achieved, independently of both the (finite) number $r$ of
hypotheses and
the special configuration of the corresponding states.
%
%th3 #&#
\begin{theorem}
\label{thmexp-decay}Let $\Sigma$ be a finite set of states on
$\mathbb{C}
^{d}$. Then there exists a~sequence $\{E_{(n)}\}_{n\in\mathbb{N}}$ of
quantum detectors relative to $\Sigma^{\otimes n}$, respectively, such
that
\[
\liminf_{n\rightarrow\infty}-\frac{1}{n}\log\mathrm{Err}_{n}\bigl(
E_{(n)}\bigr) \geq\frac{1}{3}\xi_{\mathrm{QCB}}(\Sigma).
\]
\end{theorem}

Our results are constructive in the sense that we provide an
explicitly computable quantum detector attaining the bounds.
This\vadjust{\goodbreak}
detector reduces to classical maximum likelihood in the commuting case
(cf. Section \ref{seccommute}), as does the Bayes rule, and hence
attains the optimal rate exponent (\ref{class-multip-Chern});
cf.~\cite{Salikh-73}. Thus our method can be seen as an alternative to
the quantum Bayes rule. The above error bound is a fortiori true for
the latter, and also for computable approximations to it having at
most $2$ times its error probability (Tyson
\cite{Tyson,Tyson-10}). Our results along with those of
\cite{exp-decay} allow the conjecture that in Theorem \ref{thmexp-decay}, the factor
$1/3$ can be removed; cf. also the discussion point 5 below.

To further discuss the context of the main results, we note the
following points.

1. \textit{The quantum Bayes rule} (Holevo \cite{Holevo-JMA-73}, Yuen
et al. \cite{Yuen}; cf. also Parthasa\-rathy
\cite{parth-extr,parth-ml} and Hayashi \cite{buch-Hayashi}).
Let $\Sigma=\{ \rho_{1},\ldots,\rho_{r}\}$ be such
that all
$\rho_{i}$ are distinct states on $\mathbb{C}^{d}$. Let be
$\mathcal{E}$ the set of all pertaining detectors $E$, that is,
$E=\{ E_{1},\ldots,E_{r}\} $ where $E_{i}$ are positive
self-adjoint $d\times d$ with $\sum_{i=1}^{r}E_{i}=\mathbf{1}$. Define
%
%e2.4 #&#
\begin{equation}\label{max-success}
\mu=\max_{E\in\mathcal{E}}\operatorname{Succ}(E):=\max_{E\in\mathcal{E}}
\sum_{i=1}^{r}
\mathrm{tr}[\rho_{i}E_{i}].
\end{equation}
Then there exists a unique operator $M$ on $\mathbb{C}^{d}$ satisfying
\[
\mathrm{tr}[M]=\mu,\qquad  M\geq\rho_{i},\qquad i=1,\ldots,r.
\]
Maximizers $E^{\ast}=\{ E_{1}^{\ast},\ldots,E_{r}^{\ast
}\}
\in\mathcal{E}$ of (\ref{max-success}) exist by compactness and continuity,
and any such maximizer (a Bayes rule) satisfies
%
%e2.5 #&#
\begin{eqnarray}\label{bayes-rule-descrip}
M & = & \sum_{i=1}^{r}\rho_{i}E_{i}^{\ast}=\sum_{i=1}^{r}E_{i}^{\ast
}\rho
_{i},\nonumber\\[-8pt]\\[-8pt]
( M-\rho_{i}) E_{i}^{\ast} & = & E_{i}^{\ast}(
M-\rho
_{i}) =0,\qquad i=1,\ldots,r.\nonumber
\end{eqnarray}
A proof using only elementary calculus can be found in \cite{parth-extr},
Theorem 3.1. If $r=2$, then the Holevo--Helstrom rule $\{
\mathbf{1}-\Pi,\Pi\} $ for $\Pi=\operatorname{supp}(\rho
_{2}-\rho_{1})_{+}$
is a~Bayes rule. If all states $\rho_{i}$, $i=1,\ldots,r$ commute,
hence $\rho_{i}$ can be represented as diagonal matrix with diagonal
elements $p_{ij}$, $j=1,\ldots,d$, then $M$ is a~diagonal matrix with
diagonal elements $m_{j}=\max_{i=1,\ldots,r}p_{ij}$. Then any Bayes
rule $E^{\ast}$ with diagonal matrices $E_{i}^{\ast}$ is maximum
likelihood, assigning~$0$ or $1$ to the diagonals of~$E_{i}^{\ast}$,
such that a $1$ is at $(j,j)$ only if $p_{ij}=m_{j}$.

2. \textit{Pretty good measurement.}
Let $\Sigma=\{\rho_1, \ldots, \rho_r \}$ be a set of pairwise distinct
density operators with respective a priori probabilities $p_i$. Define
the positive semi-definite operator $\rho=\sum_{i=1}^r p_i \rho_i$. A
possible quantum detector relative to $\Sigma$ is of the form
\[
E_i^{\mathrm{PGM}}:= \rho^{-1/2}p_i \rho_i \rho^{-1/2}, \qquad i=1,\ldots,r.
\]
(The inverse is understood to be taken on the support of $\rho$ only.)
It represents the widely investigated POVMs called pretty good
measurements (PGM). These are known to be a good approximation of the
quantum Bayes rule: if~$\Sigma$ is a set of pure states, then the
averaged success probability $\operatorname{Succ}(\mathrm{PGM})=\sum
_{i=1}^r p_i \operatorname{Succ}_i(\mathrm{PGM})$ is lower bounded by a
result of Barnum and Knill \cite{BarnumKnill},
\[
\operatorname{Succ}(\operatorname{PGM})\geq
\Biggl( \max_{E\in\mathcal{E}}\sum_{i=1}^r p_i
\operatorname{Succ}_i(E)\Biggr)^2,
\]
where $\mathcal{E}$ denotes the set of quantum detectors relative to
$\Sigma$. For further bounds on $\operatorname{Succ}(\mathrm{PGM})$ refering
also to the general case of mixed states see~\cite{Montan} and
references therein. To the best of our knowledge, in the literature,
the PGM has not been successfully used to study the optimal asymptotic
error exponent.

3. \textit{Classical results resembling the multiple Chernoff bound.}
Let $\Sigma$ be a~statistical experiment having finite parameter space
$\{ \theta_{1},\ldots,\theta_{r}\} $, and $\Sigma^{n}$ be
the associated product experiment corresponding to i.i.d.
observations. Torgersen \cite{Torg} considered $\delta( \Sigma
^{n},\Sigma_{a}) $, the deficiency (in the Le Cam sense) of~$\Sigma^{n}$ with respect to the fully informative experiment
$\Sigma_{a}$. Here $\Sigma_{a}$ may be identified, up to equivalence,
with the set of $r$ point masses concentrated on~$\theta_{1},\ldots,\theta_{r}$. It was shown (\cite{Torg}, Theorem
4.2) that
\[
-\frac{1}{n}\log\delta( \Sigma^{n},\Sigma_{a})
\rightarrow
\xi_{\mathrm{CB}}(\Sigma)\qquad\mbox{as }n\rightarrow\infty
\]
with $\xi_{\mathrm{CB}}(\Sigma)$ defined in (\ref{class-CB-def}). Krob and von
Weizs\"{a}cker \cite{Krob-vonWeiz} considered the Shannon capacity
$C(\Sigma^{n})$ of $\Sigma^{n}$ construed as a communication channel,
and showed that $C(\Sigma^{n})$ approaches its upper bound $\log r$
exponentially quickly, with rate exponent $\xi_{\mathrm{CB}}(\Sigma)$:
\[
-\frac{1}{n}\log\bigl( \log r-C(\Sigma^{n})\bigr) \rightarrow\xi
_{\mathrm{CB}}(\Sigma)\qquad\mbox{as }n\rightarrow\infty.
\]

4. \textit{Linearly independent states.}
A stronger condition than Condition \ref{condLI} would be that all states $\{
\rho_{1},\ldots,\rho_{r}\} $ are linearly independent (in the
sense that for any selected $r$ bases of the spaces $\operatorname{supp}
(\rho_{i}),i=1,\ldots,r$, the union set of vectors is linearly
independent.) The paper \cite{eldar} gives examples of such ensembles
of states, and shows that under this stronger condition, the Bayes
detector $E=\{ E_{1},\ldots,E_{r}\} $ consists of
projections $E_{i}$ (is a von Neumann measurement or PVM). Lemma
\ref{lemasymp-norm-of-Gram-matrices} implies that our pairwise
Condition \ref{condLI}
on $\Sigma$ implies the stronger one for $\Sigma^{\otimes n}$,
that is, the states $\rho_{1}^{\otimes},\ldots,\rho_{r}^{\otimes}$ are
linearly independent for sufficiently large $n$.

5. \textit{Other special ensembles.}
It can be shown that there are other situations besides Condition \ref{condLI}
where the error exponent $\xi_{\mathrm{QCB}}( \Sigma) $ is
attainable exactly. One condition, which does not impose any rank
restrictions on the states and thus allows for full rank density
matrices $\rho_{i}$, is as follows. For a~set $\Sigma=\{
\rho_{1},\ldots,\rho_{r}\} $ of density operators where
$r>2$, let
$\Sigma_{(i,j)-}$ be the set where a pair $\rho_{i}$, $\rho_{j}$, is
removed, that is, $\Sigma_{(i,j)-}=\Sigma\setminus\{
\rho_{i},\rho_{j}\} $ for $1\leq i<j\leq r$. Assume there is a
pair $(i,j)$ such that
\[
\xi_{\mathrm{QCB}}( \Sigma) \leq\tfrac{1}{6}\xi_{\mathrm{QCB}}\bigl(
\Sigma
_{(i,j)-}\bigr) .%\label{new-condition}%
\]
This condition can replace Condition \ref{condLI} in the statement of Theorem
\ref{thmmultiple-Chernoff}, that is, the multiple quantum Chernoff
bound is then attainable. The proof, not to be presented here,
consists in a combination of the sample splitting method of
\cite{exp-decay} with Theorem
\ref{thmexp-decay}. This further supports the conjecture that the
result of Theorem \ref{thmexp-decay} is not final and the factor
$1/3$ there may be removed.

Throughout the paper, we use the notation $j\in\{ 1,\ldots,d\}
$ and $j\in[ 1,d] $ interchangeably.

%s3 #&#
\section{The detection algorithm}\label{secalgorithm}

In this section, we construct a sequence~$E^{(n)}$, $n\in\mathbb{N}$,
of quantum detectors for $\Sigma^{\otimes n}$. The construction does
not rely on the existence of asymptotically optimal quantum tests for
the binary case. It is rather a modification of a construction used
in \cite{NSz2} which yields asymptotically optimal quantum tests for a
set of pure states. At the same time, it represents a quantum
extension of the classical ML method, different from the Bayes rule
described in~(\ref{bayes-rule-descrip}).

Consider again the classical case where a set $\Sigma=\{ P_{1},
\ldots,P_{r}\} $ of probability distributions is given on a
finite sample space $\Omega$ with cardinality $d$. An obvious
algorithmic description of a ML decision rule
$\varphi\dvtx\Omega\rightarrow\{ 1,\ldots,r\} $ is as
follows. For each $\omega\in\Omega$, find a maximal element in
$\{ P_{i}(\omega)\} _{i=1}^{r}$, say~$P_{i^{\ast}}(\omega)$, and then decide
$\varphi(\omega)=i^{\ast}$. Alternatively, one may successively find
the largest probabilities among all $P_{i}(\omega)$, identify which
$P_{i}$ and which $\omega$ they are from, and assign a corresponding
decision on this $\omega$. This iterative approach can be expressed in
a simple algorithm in pseudocode as follows.
%
%al1 #&#
\begin{algorithm}[(Classical ML rule)]
\label{alg-ML}

\textit{Initialize.} Let
$\Pi_{0}=\{
P_{i}(\omega), i=1,\ldots,r, \omega\in\Omega\} $ be the
$r\times d$-matrix of all probabilities.

\textit{For $s=1$ to $d$}:

\begin{longlist}
\item
In $\Pi_{s-1}$ find a maximal entry,
$P_{i^{\ast}}(\omega^{\ast})$ say. Set $\omega_{s}=\omega^{\ast}$ and
decide $\varphi( \omega_{s}) =i^{\ast}$.

\item
In $\Pi_{s-1}$, all $P_{i} (\omega_{s}), i=1,\ldots,r$ are replaced by
$-1$; the resulting $r\times d$-matrix is $\Pi_{s}$.
\end{longlist}
\end{algorithm}

After $s=d$ steps, the matrix $\Pi_{s}$ has entries $-1$ only (a value
serving as an indicator, chosen to be smaller than any
probability). We also have enumerated the elements of $\Omega$ as
$\omega_{1},\ldots,\omega_{d}$; on each of these, a decision~$\varphi( \omega_{s}) $ has been made, which is ML by
construction.

In the quantum case, there is no initial sample space $\Omega$; it only
appears after defining a \textit{measurement}, which in our
context\vadjust{\goodbreak}
can be
taken to be an orthonormal basis $\{ e_{s}\} _{s=1}^{d}$ of
$\mathbb{C}^{d}$. After this basis is fixed, the sample space $\Omega
=\{
\omega_{s}\} _{s=1}^{d}$ can be identified with the basis
itself, or
more precisely with the set of pertaining projectors, such that each
$\omega_{s}=|e_{s}\rangle\langle e_{s}|$, and a classical nonrandomized
decision rule $\varphi\dvtx\Omega\rightarrow\{ 1,\ldots,r
\} $ has to
be found. Then the quantum decision rule $E=\{ E_{1},\ldots
,E_{r}\} $ is given by the PVM
%
%e3.1 #&#
\begin{equation} \label{pvm-construct}
E_{i}=\sum_{s\dvtx\varphi( |e_{s}\rangle\langle e_{s}|) =i}
|e_{s}\rangle\langle e_{s}|,\qquad i=1,\ldots,r.
\end{equation}

The algorithm we will describe constructs the basis elements $e_{j}$
and the pertaining decision $\varphi( \cdot) $ iteratively,
combining the ML principle underlying Algorithm \ref{alg-ML} with a
Gram--Schmidt orthogonalization.

For each $1\leq i\leq r$ let
%
%e3.2 #&#
\begin{equation}\label{spec-decompose-rho}
\rho_{i}=\sum_{j=1}^{d}\lambda_{ij}|v_{ij}\rangle\langle v_{ij}|
\end{equation}
be a spectral decomposition of the density matrix $\rho_{i}$, where
$\lambda_{ij}$, $j=1,\ldots,d$, are the eigenvalues of $\rho_{i}$
appearing with their multiplicity, in arbitrary order, and
$|v_{ij}\rangle$ are the corresponding normalized eigenvectors in
$\mathbb{C}^{d}$. Here~$\langle v_{ij}|$ denotes the dual vector such
that in this notation $|v_{ij} \rangle\langle v_{ij}|$ describes an
orthogonal projector onto the one-dimensional subspace of
$\mathbb{C}^{d}$ spanned by~$|v_{ij}\rangle$. We stress that zero
eigenvalues are included with their multiplicity since $d$ in
(\ref{spec-decompose-rho}) is the dimension of $\rho_{i}$.
%
%al2 #&#
\begin{algorithm}[(A quantum decision rule)]
\label{alg-GS}

\textit{Initialize.}
Let $\Lambda_{0}=\{ \lambda_{ij}, i=1,\ldots,r, j=1,\ldots
,d\}
$ be the $r\times d$-matrix of all eigenvalues. Let $e_{0}=0$ be the zero
vector in $\mathbb{C}^{d}$.

\textit{For $s=1$ to $d$}:
\begin{longlist}
\item
In $\Lambda_{s-1}$ find a maximal entry,
$\lambda_{i^{\ast}j^{\ast}}$ say. Set $e_{s}$ to be a unit vector such
that
%
%e3.3 #&#
\begin{equation} \label{G-S-orthogonalize-crucial}
e_{s}\in\operatorname{span}(e_{1},\ldots,e_{s-1},v_{i^{\ast}j^{\ast}}),\qquad
e_{s}\perp\operatorname{span}(e_{1},\ldots,e_{s-1})
\end{equation}
and decide $\varphi( |e_{s}\rangle\langle e_{s}|)
=i^{\ast}
$.

\item In $\Lambda_{s-1}$, all $\lambda_{ij}$ such
that $v_{ij}\in\operatorname{span}(e_{1},\ldots,e_{s})$ are replaced by
$-1$; the resulting $r\times d$-matrix is $\Lambda_{s}$.
\end{longlist}
\end{algorithm}

Again, after $s=d$ steps, the matrix $\Lambda_{s}$ has entries $-1$
only. We also have constructed an orthonormal basis
$e_{1},\ldots,e_{d}$ and on each of these, an associated decision
$\varphi( |e_{s}\rangle\langle e_{s}|) $. The crucial step
(\ref{G-S-orthogonalize-crucial}) is recognized to define a
Gram--Schmidt orthogonalization process. The quantum detector now is
given by the PVM (\ref{pvm-construct}).

To bound the error probability of this detector, we need to introduce
some further notation. In each step $s$ of Algorithm \ref{alg-GS}, in
part
(i) we have selected an index pair $( i^{\ast},j^{\ast
})
$ where $\lambda_{i^{\ast}j^{\ast}}$ is a maximal entry of the matrix
$\Lambda_{s-1}$; set $( i(s),j(s)) =$ $(
i^{\ast},j^{\ast}) $. The sequence of vectors $\{
v_{i(s),j(s)}\} _{s=1}^{d}$ is linearly independent by
construction. For each $s\in[ 1,d] $ define a
$d\times s$ matrix $V_{s}$
%
%e3.4 #&#
\begin{equation} \label{Vs-def}
V_{s}:=\bigl(v_{i(1),j(1)},\ldots,v_{i(s),j(s)}\bigr),
\end{equation}
that is, the columns of $V_{s}$ are the vectors $v_{i(k),j(k)}$,
$k\in[ 1,s] $. We refer to the $s\times s$-matrix
%
%e3.5 #&#
\begin{equation} \label{Gamma-s-def}
\Gamma_{s}:=V_{s}^{\ast}V_{s}
\end{equation}
as a Gram matrix of $\{ v_{i(k),j(k)}\} _{k=1}^{s}$. For
each $s\in[1,d]$ the matrix $\Gamma_{s}$ is nonsingular and the
matrix
\[
P_{s}:=V_{s}(V_{s}^{\ast}V_{s})^{-1}V_{s}^{\ast}=V_{s}\Gamma_{s}^{-1}
V_{s}^{\ast}
\]
represents\vspace*{1pt} an orthogonal projection onto $\operatorname{span}(v_{i(1),j(1)}
,\ldots,v_{i(s),j(s)})$, an $s$-di\-mensional subspace of
$\mathbb{C}^{d}$. Additionally, we set $P_{0}=0$ and define for
$s\in[1,d]$
%
%e3.6 #&#
\begin{equation} \label{Ps-def}
P^{(s)}:=P_{s}-P_{s-1}.
\end{equation}
Observe that the $P^{(s)}$ represent one-dimensional orthogonal
projectors, which are mutually orthogonal, such that
$P^{(s)}=|e_{s}\rangle\langle e_{s}|$ for the unit vectors $e_{s}$
defined in (\ref{G-S-orthogonalize-crucial}). The latter
can be taken to be $e_{s}=\Vert P^{(s)}v_{i(s),j(s)}
\Vert^{-1}P^{(s)}v_{i(s),j(s)}$ (or a sign changed version).

Furthermore, define an index $N$ as
%
%e3.7 #&#
\begin{equation} \label{N-def}
N=\max\bigl\{ s\in[1,d]\dvtx\lambda_{i(s),j(s)}>0\bigr\}.
\end{equation}
It can be seen from the proof of Lemma \ref{lem-general-algorithm}
below that if $N<d$, then $N$ can serve as an early stopping index for
Algorithm \ref{alg-GS}, in the following sense: the obtained set of
orthonormal vectors $\{ e_{s}\} _{s=1}^{N}$ can be
completed to a basis of $\mathbb{C}^{d}$ in an arbitrary way and the
decisions $\varphi( e_{s}) $, $s>N$, can be taken
arbitrarily. This is related to the fact that for all further steps
$s>N$, the remaining eigenvalues $\lambda_{ij}$ listed in the matrix
$\Lambda_{s}$ are $0$; in Algorithm \ref{alg-ML} this corresponds to
the case that there exist $\omega\in\Omega$ which are outside the
support of all $P_{i}$.

We use the notation $\lambda_{\min}(\cdot)$ for the minimal eigenvalue of
a self-adjoint matrix.
%
%le1 #&#
\begin{lemma}
\label{lem-general-algorithm}Let $\Sigma=\{\rho_{i}\}_{i=1}^{r}$ be an
arbitrary set of density matrices on~$\mathbb{C}^{d}$. Then the
detector $E=\{E_{i}\}_{i=1}^{r}$ constructed in Algorithm \ref{alg-GS}
fulfills
%
%e3.8 #&#
\begin{equation}\label{estav-err}
\operatorname{Err}(E)\leq\lambda_{\min}^{-1}(\Gamma_{N}) r^{-1}\sum
_{1\leq i,j\leq
r,j\not=i}\inf_{s\in[0,1]}\operatorname{tr}
[\rho_{i}^{1-s}\rho_{j}^{s}],
\end{equation}
where $\Gamma_{N}$ is the Gram matrix according to (\ref{Gamma-s-def})
for index $s=N$ defined in~(\ref{N-def}).
\end{lemma}
\begin{pf}
Define $J$ to be the subset of $[ 1,r] \times[ 1,d] $ consisting of
all
pairs $( i(s),j(s))$, $s\in[ 1,d] $,\vadjust{\goodbreak} and $J_{i}:=\{ j\in[ 1,d] \dvtx (
i,j) \in J\} $. For given $i\in[1,r]$, consider the corresponding
individual success probability of the detector defined
by~(\ref{pvm-construct}):
%
%e3.9 #&#
\begin{equation}\label{first-low-bound}
\mathrm{Succ}_{i}(E)=\operatorname{tr}[\rho_{i}E_{i}]=\sum
_{j=1}^{d}\lambda_{ij}\operatorname{tr}[|v_{ij}\rangle\langle v_{ij}
|E_{i}]\geq\sum_{j\in J_{i}}\lambda_{ij}\operatorname{tr}[|v_{ij}
\rangle\langle v_{ij}|E_{i}],\hspace*{-28pt}
\end{equation}
where the right-hand side is set $0$ if the set $J_{i}$ is empty. For
any $j\in J_{i}$, let~$s(i,j)$ be the unique index $s\in[1,d]$
such that $(i,j)=( i(s),j(s)) $. If $J_{i}$ is nonempty,
then
\[
E_{i}=\sum_{j\in J_{i}}\bigl|e_{s(i,j)}\bigr\rangle\bigl\langle
e_{s(i,j)}\bigr|=\sum_{j\in J_{i} }P^{(s(i,j))}
\]
with $P^{(s)}$ defined in (\ref{Ps-def}), hence $E_{i}\geq
P^{(s(i,j))}$ for all $j\in J_{i}$, in the sense of the ordering for
self-adjoint matrices. This implies
\[
\operatorname{tr}[|v_{ij}\rangle\langle v_{ij}|E_{i}]\geq\bigl\langle
v_{ij}\bigl|P^{(s(i,j))}\bigr|v_{ij}\bigr\rangle=\bigl\langle
v_{ij}\bigl|P_{s(i,j)} \bigr|v_{ij}\bigr\rangle-\bigl\langle
v_{ij}\bigl|P_{s(i,j)-1}\bigr|v_{ij}\bigr\rangle.
\]
Recall that the matrices $P_{s}$ are constructed as orthogonal
projectors onto
$\operatorname{span}(v_{i(1),j(1)},\ldots,v_{i(s),j(s)})$, and since for
$j\in J_{i}$ and $s=s(i,j)$ we have $v_{ij}=v_{i(s),j(s)}$, it follows
that for $s=s(i,j)$
\[
\bigl\langle v_{ij}\bigl|P_{s(i,j)}\bigr|v_{ij}\bigr\rangle=\bigl\langle
v_{i(s),j(s)}|P_{s}|v_{i(s),j(s)}\bigr\rangle=1.
\]
Consequently,
\[
\mathrm{Succ}_{i}(E)\geq\sum_{j\in J_{i}}\lambda_{ij}\bigl\langle
v_{ij}\bigl|P^{(s(i,j))}\bigr|v_{ij}\bigr\rangle=\sum_{j\in J_{i}}\lambda_{ij}
-\sum_{j\in J_{i}}\lambda_{ij}\bigl\langle v_{ij}\bigl|P_{s(i,j)-1}\bigr|v_{ij}
\bigr\rangle.
\]
For the individual error probability under state $\rho_{i}$ this
implies, setting $J_{i}^{c}:=[ 1$, $d] \setminus J_{i}$,
%
%e3.10 #&#
\begin{eqnarray}\label{err-bound-projec-a}
\mathrm{Err}_{i}(E) &=& 1-\mathrm{Succ}_{i}(E)\leq\sum_{j\in J_{i}}
\lambda_{ij}\bigl\langle v_{ij}\bigl|P_{s(i,j)-1}\bigr|v_{ij}\bigr\rangle
+\sum_{j\in
J_{i}^{c}}\lambda_{ij}\nonumber\\[-8pt]\\[-8pt]
&=& S_{1}+S_{2},\nonumber
\end{eqnarray}
say.

\textit{Bounding the term} $S_{1}$. Consider only those terms in
\[
S_{1}=\sum_{j\in J_{i}}\lambda_{ij}\bigl\langle v_{ij}\bigl|P_{s(i,j)-1}
\bigr|v_{ij}\bigr\rangle,
\]
where $\lambda_{ij}>0$. Since for $j\in J_{i}$ we have $\lambda_{ij}
=\lambda_{i(s),j(s)}$ for some $s=s(i,j)\in[ 1,d] $, the
assumption $\lambda_{ij}>0$ implies $s(i,j)\leq N$. Recall that $P_{s}
=V_{s}\Gamma_{s}^{-1}V_{s}^{\ast}$, $s=1,\ldots,d$, and that each
$\Gamma_{s-1}$ is a principal submatrix of $\Gamma_{s}$. As a
consequence,
$\lambda_{\min}(\Gamma_{s})\geq\lambda_{\min}(\Gamma_{N})$,
$s\in[ 1,N] $, and for $j\in J_{i}$, if not $s(i,j)=1$,
%
%e3.11 #&#
\begin{eqnarray}\label{principal-sub}\quad
\lambda_{ij}\bigl\langle v_{ij}\bigl|P_{s(i,j)-1}\bigr|v_{ij}\bigr\rangle
&\leq& \lambda_{\min}^{-1}\bigl(\Gamma_{s(i,j)-1}\bigr)\lambda_{ij}\bigl\langle v_{ij}
\bigl|V_{s(i,j)-1}V_{s(i,j)-1}^{\ast}\bigr|v_{ij}\bigr\rangle\nonumber\\[-8pt]\\[-8pt]
&\leq& \lambda_{\min}^{-1}(\Gamma_{N})\lambda_{ij}\bigl\langle v_{ij}
\bigl|V_{s(i,j)-1}V_{s(i,j)-1}^{\ast}\bigr|v_{ij}\bigr\rangle,\nonumber
\end{eqnarray}
where $\lambda_{\min}(\Gamma_{N})>0$ by construction. Formally setting
$V_{0}=0\in\mathbb{C}^{d}$, the above inequality holds also if
$s(i,j)=1$. One obtains the upper bound
%
%e3.12 #&#
\begin{eqnarray}\label{estind-err-0-a}
S_{1} & = & \sum_{j\in J_{i}}\lambda_{ij}\bigl\langle v_{ij}\bigl|P_{s(i,j)-1}
\bigr|v_{ij}\bigr\rangle\nonumber\\
&\leq&\lambda_{\min}^{-1}(\Gamma_{N})\sum
_{j\in
J_{i} }\lambda_{ij}\bigl\langle
v_{ij}\bigl|V_{s(i,j)-1}V_{s(i,j)-1}^{\ast}\bigr|v_{ij}
\bigr\rangle\\
& = & \lambda_{\min}^{-1}(\Gamma_{N})\sum_{j\in J_{i}}\lambda
_{ij}\sum
_{k=1}^{s(i,j)-1}\bigl|\bigl\langle v_{i(k),j(k)}|v_{ij}\bigr\rangle\bigr|^{2}.
\nonumber
\end{eqnarray}
The identity above is based on the fact that the columns of
$V_{s(i,j)-1}$ are given by the vectors $v_{i(k),j(k)}$,
$k\in[1,s(i,j)-1]$. Note that in (\ref{estind-err-0-a}), for every
pair of vectors occurring in $\langle v_{i(k),j(k)}|v_{ij}\rangle$
the corresponding eigenvalues satisfy $\lambda_{i(k),j(k)}\geq\lambda
_{ij}$ by construction. This implies
%
%e3.13 #&#
\begin{equation}\label{inequ-eigenval-a}
\lambda_{ij}\leq\lambda_{ij}^{1-s}\lambda_{i(k),j(k)}^{s}
\end{equation}
for every $s\in[0,1]$. Recall that every eigenvalue
$\lambda_{i(k),j(k)}$ pertains to a sta\-te~$\rho_{i(k)}$; we may
assume $i(k)\neq i$, since otherwise necessarily $j(k)\neq j$ and thus
$\langle v_{i(k),j(k)} |v_{i(k),j}\rangle=0$. Setting now $m=i(k)$ and
assuming $m\neq i$, we will apply inequality (\ref{inequ-eigenval-a})
for an exponent $s$ which is allowed to depend on $i$ and $m$. Denote
by $s(i,m)=$ $s(m,i)$ $\in[0,1]$ the exponent associated to the
pair of indices $( i,m) \in[ 1,r] ^{2}$. Observe
that for any subset $D_{m}\subset[ 1,d] $
%
%e3.14 #&#
\begin{eqnarray}\label{inequ-sum-eigenval-b}
&&
\sum_{j\in J_{i}}\sum_{j^{\prime}\in D_{m}}\lambda_{ij}^{1-s(i,m)}
\lambda_{m,j^{\prime}}^{s(i,m)}|\langle v_{m,j^{\prime
}}|v_{ij}\rangle
|^{2}\nonumber\\[-8pt]\\[-8pt]
&&\qquad\leq\sum_{j\in J_{i}}\sum_{j^{\prime}=1}^{d}\lambda_{ij}^{1-s(i,m)}
\lambda_{m,j^{\prime}}^{s(i,m)}|\langle v_{m,j^{\prime
}}|v_{ij}\rangle|^{2},\nonumber
\end{eqnarray}
where on the right-hand side of the inequality we are just adding
positive reals. It now follows from (\ref{estind-err-0-a}),
(\ref{inequ-eigenval-a}) and (\ref{inequ-sum-eigenval-b}) that
%
%e3.15 #&#
\begin{equation}\label{est-ind-err-5-a}
S_{1}\leq\lambda_{\min}^{-1}(\Gamma_{N})\sum_{j\in J_{i}}\sum
_{1\leq
m\leq r,m\not
=i}\sum_{j^{\prime}=1}^{d}\lambda_{ij}^{1-s(i,m)}\lambda
_{m,j^{\prime
}}^{s(i,m)}|\langle v_{m,j^{\prime}}|v_{ij}\rangle|^{2}.
\end{equation}

\textit{Bounding the term} $S_{2}$. We have
\[
S_{2}=\sum_{j\in J_{i}^{c}}\lambda_{ij}=\sum_{j\in J_{i}^{c}}\lambda
_{ij}\langle v_{ij}|v_{ij}\rangle.
\]
Consider only those terms where $\lambda_{ij}>0$. By definition of $J_{i}^{c}
$, there exists $s\in[1,d]$ such that $v_{ij}\in\operatorname{span}
(v_{i(1),j(1)},\ldots,v_{i(s),j(s)})$. Then $\lambda_{i(k),j(k)}\geq
\lambda_{ij}$ for $k\in[1,s]$, hence $\lambda_{i(s),j(s)}>0$ and
consequently $s\leq N$. We also have $\langle v_{ij}|v_{ij}\rangle
=\langle
v_{ij}|P_{s}|v_{ij}\rangle$, so the same reasoning as for $S_{1}$
leads to
%
%e3.16 #&#
\begin{equation}\label{est-ind-err-6-a}
S_{2}\leq\lambda_{\min}^{-1}(\Gamma_{N})\sum_{j\in
J_{i}^{c}}\sum_{1\leq m\leq r,m\not
=i}\sum_{j^{\prime}=1}^{d}\lambda_{ij}^{1-s(i,m)}\lambda
_{m,j^{\prime
}}^{s(i,m)}|\langle v_{m,j^{\prime}}|v_{ij}\rangle|^{2}.
\end{equation}
Putting together (\ref{est-ind-err-5-a}) and (\ref{est-ind-err-6-a}),
we obtain
\[
\mathrm{Err}_{i}(E)\leq\lambda_{\min}^{-1}(\Gamma_{N})\sum_{j=1}^{d}
\sum_{1\leq m\leq r,m\not=i}\sum_{j^{\prime}=1}^{d}\lambda_{ij}
^{1-s(i,m)}\lambda_{m,j^{\prime}}^{s(i,m)}|\langle v_{m,j^{\prime}}
|v_{ij}\rangle|^{2}.
\]
Since $s(i,m)$, $m\not=i$, are arbitrary in $[0,1]$, we obtain
\[
\mathrm{Err}_{i}(E)\leq\lambda_{\min}^{-1}(\Gamma_{N})\sum_{1\leq
m\leq
r,m\not=i}\inf_{s\in[0,1]}\operatorname{tr}
[\rho_{i}^{1-s}\rho_{m}^{s}].
\]
By averaging over $i\in[ 1,r] $, we obtain
(\ref{estav-err}).
\end{pf}

%s4 #&#
\section{Pairwise linearly independent states}%\label{seclinindep}
\label{secachievability}

The main difficulty for utilizing Lemma \ref{lem-general-algorithm}
for an asymptotic error bound is the control of the minimal eigenvalue
of the Gram matrix $\Gamma_{N}$. Imposing Condition \ref{condLI} on the set
$\Sigma=\{ \rho_{1},\ldots,\rho_{r}\}$ is one way to
achieve that control, resulting in Theorem~\ref{thmmultiple-Chernoff}.
Observe that this condition is equivalent
to requiring that for each pair $\rho_{i},\rho_{j}$, $i\neq j$, the
joint set of eigenvectors pertaining to a nonzero eigenvalue is
linearly independent. Lemma \ref{lemasymp-norm-of-Gram-matrices}
below implies in this case: the Gram matrix~$\Gamma_{N}$ associated to
the tensor product set $\Sigma^{\otimes n}=\{
\rho_{1}^{\otimes n},\ldots,\rho_{r}^{\otimes n}\} $ has minimal
eigenvalue bounded away from zero as $n\rightarrow\infty$.

For each of the original $\rho_{i}$, let
$d_{i}:=\operatorname{rank}(\rho_{i})$ the number of nonzero
eigenvalues. Condition \ref{condLI} implies that for any $i\neq j$ we have
$d_{i}+d_{j}\leq d$, and since $d_{i}\geq1$ this implies that all
$d_{i}<d$. In this case
$\operatorname{rank}(\rho_{i}^{\otimes n})=d_{i}^{n}<d^{n}$. Let
$\mathcal{V}_{n}$ be the set of eigenvectors of $\rho_{1}^{\otimes n}
,\ldots,\rho_{r}^{\otimes n}$ pertaining to a nonzero eigenvalue; more
precisely, if we assume spectral representations
\[
\rho_{i}^{\otimes n}=\sum_{j=1}^{\mathrm{rank}(\rho_{i}^{\otimes
n})}\lambda
_{ij}|v_{ij}\rangle\langle v_{ij}|
\]
with unit vectors $v_{ij}$ and eigenvalues $\lambda_{ij}>0$, then
$\mathcal{V}_{n}$ is the double array
\[
\mathcal{V}_{n}=\{ v_{ij},j\in[ 1,d_{i}^{n}],
i\in[
1,r] \}
\]
so that $\#\mathcal{V}_{n}=D_{n}:=\sum_{i=1}^{r}d_{i}^{n}$.
%
%le2 #&#
\begin{lemma}
\label{lemasymp-norm-of-Gram-matrices} Let $\Sigma=\{ \rho_{1}
,\ldots,\rho_{r}\} $ be a set of density matrices in
$\mathbb{C}^{d}$, fulfilling Condition \ref{condLI}. Let $\mathcal{V}_{n}$ be
the set of eigenvectors defined above and let $\mathring{\Gamma}_{n}$
its $D_{n}\times D_{n}$ Gram matrix. Then
%
%e4.1 #&#
\begin{equation} \label{idnlemma4}
\lambda_{\min}( \mathring{\Gamma}_{n}) =1+o(1)\qquad\mbox{as
}n\rightarrow\infty.
\end{equation}
\end{lemma}
\begin{pf}
We will first argue for the generic case $n=1$, and subsequently
impose the tensor product structure on the $\rho_{i}$. As above, let
$\{ v_{ij}\} _{j=1}^{d_{i}}$ be the eigenvectors of
$\rho_{i}$ pertaining to a nonzero eigenvalue. Define a $d\times
d_{i}$ matrix
%
%e4.2 #&#
\begin{equation} \label{U-i-def-a}
U_{i}:=(u_{i1},\ldots,u_{id_{i}}),
\end{equation}
that is, the columns of $U_{i}$ are the vectors $u_{ij}$, $j\in[
1,d_{i}] $. Furthermore, define a $d\times D$ matrix (where
$D=\sum_{i=1}^{r}d_{i}$)
\[
U:=(U_{1}|\cdots|U_{r})
\]
made up of submatrices $U_{i}$. Now, for $n>1$ replace the matrices
$U_{i}$ in (\ref{U-i-def-a}) by their $n$th tensor powers
$U_{i}^{\otimes n}$. Then for $n\geq1$ the $d^{n}\times d_{i}^{n}$
blocks $U_{i}^{\otimes n}$ correspond to eigenvectors of
$\rho_{i}^{\otimes n}$, and $U$ is now of dimension $d^{n}\times
D_{n}$ where $D_{n}=\sum_{i=1}^{r}d_{i}^{n}$. For the $D_{n}\times
D_{n}$ Gram matrix $\mathring{\Gamma}_{n}:=U^{\ast}U$ we show
(\ref{idnlemma4}).

We will again begin with the case $n=1$ and develop a representation
of~$U$ which takes account of its block structure in terms of
$U_{i}^{\ast}U_{j}$. To this end, for $i\in[ 1,r] $ define
$d_{i}\times D$ matrices
\[
E_{i}=(0_{d_{i}\times d_{1}}|\cdots|0_{d_{i}\times d_{i-1}}|\mathbf{1}_{d_{i}
}|0_{d_{i}\times d_{i+1}}|\cdots|0_{d_{i}\times d_{r}}),
\]
where we denote a $k\times l$ matrix of $0$'s by $0_{k\times l}$ and the
$k$-dimensional unit matrix by $\mathbf{1}_{k}$. Then it is easily
seen that
$U=\sum_{i=1}^{r}U_{i}E_{i}$ and consequently
%
%e4.3 #&#
\begin{equation} \label{Gamma-1-decomp-a}
\mathring{\Gamma}_{1}=U^{\ast}U=\sum_{i,j=1}^{r}E_{i}^{\ast
}U_{i}^{\ast}
U_{j}E_{j}.
\end{equation}
Here $U_{i}^{\ast}U_{i}=\mathbf{1}_{d_{i}}$, $i\in[ 1,r
] $,
so that
\[
{\mathbf{1}}_{D}=\sum_{i=1}^{r}E_{i}^{\ast}U_{i}^{\ast}U_{i}E_{i}.
\]
We define
%
%e4.4 #&#
\begin{equation} \label{delta-def-a}
\Delta:=\mathring{\Gamma}_{1}-{\mathbf{1}}_{D}
\end{equation}
and write $\mathring{\Gamma}_{1}={\mathbf{1}}_{D}+\Delta$. Moreover,
for $j<i$ we define
%
%e4.5 #&#
\begin{equation} \label{delta-ij-def-a}
\Delta_{ij}=E_{i}^{\ast}U_{i}^{\ast}U_{j}E_{j}+E_{j}^{\ast
}U_{j}^{\ast}
U_{i}E_{i}.
\end{equation}
Clearly $\Delta_{ij}$ is Hermitian, and by construction $\Delta=\sum_{i=2}
^{r}\sum_{j=1}^{i-1}\Delta_{ij}$. Now, with $\Vert a\Vert
=\lambda_{\max}^{1/2}( a^{2}) $ being the operator norm
of a
Hermitian matrix $a$, we have
%
%e4.6 #&#
\begin{eqnarray}\label{ev-estim-2b-a}
\lambda_{\min}( \mathring{\Gamma}_{1}) &=& \min_{
\Vert
v\Vert=1}\langle v|{\mathbf{1}}_{D}+\Delta|v
\rangle
=1+\min_{\Vert v\Vert=1}\langle
v|\Delta|v\rangle
\nonumber\\[-8pt]\\[-8pt]
&\geq& 1-\Vert\Delta\Vert\geq1-\sum_{i=2}^{r}\sum_{j=1}
^{i-1}\Vert\Delta_{ij}\Vert=1-\sum_{i=2}^{r}\sum_{j=1}
^{i-1}\lambda_{\max}^{1/2}( \Delta_{ij}^{2}),\nonumber
\end{eqnarray}
where the second inequality is by the triangle inequality for the
operator norm.\vadjust{\goodbreak}

For the case $n>1$, replacing the matrices $U_{i}$ in
(\ref{U-i-def-a}) by their $n$th tensor powers $U_{i}^{\otimes n}$
leads to a representation of $\mathring{\Gamma}_{n}$ analogous to
(\ref{Gamma-1-decomp-a}). Here the matrices $E_{i}$ have to be
replaced by $E_{i,n}$, defined analogously to $E_{i}$ with~$d_{i}$
replaced by $d_{i}^{n}$, $i\in[ 1,r] $. Furthermore, we
define $\Delta_{n}$ and $\Delta_{ij,n}$ analogously to~(\ref{delta-def-a})
and (\ref{delta-ij-def-a}) with $U_{i}$, $E_{i}$
replaced by $U_{i}^{\otimes n}$ and $E_{i,n}$. In order to prove~%
(\ref{idnlemma4}) we use the analog of (\ref{ev-estim-2b-a}) holding
for $\mathring{\Gamma}_{n}$ and $\Delta_{n}$, which is
\[
\lambda_{\min}( \mathring{\Gamma}_{n}) \geq1-\sum_{i=2}^{r}
\sum_{j=1}^{i-1}\lambda_{\max}^{1/2}( \Delta_{ij,n}^{2}).
\]
It now suffices to show that for all $i\in[ 2,r] $,
$j\in[ 1,i-1] $
%
%e4.7 #&#
\begin{equation} \label{delta-ij-bound-a}
\lambda_{\max}^{1/2}( \Delta_{ij,n}^{2}) \rightarrow
0\qquad\mbox{as }n\rightarrow\infty.
\end{equation}
Clearly, we have
\[
\Delta_{ij,n}=E_{i,n}^{\ast}( U_{i}^{\ast}U_{j}
)^{\otimes
n}E_{j,n}+E_{j,n}^{\ast}( U_{j}^{\ast}U_{i})^{\otimes
n}E_{i,n}
\]
and by a computation, since
$E_{i,n}E_{i,n}^{\ast}=\mathbf{1}_{d_{i}^{n}}$ and
$E_{j,n}E_{i,n}^{\ast}=0_{d_{j}^{n}\times d_{i}^{n}}$ for $j<i$,
\[
\Delta_{ij,n}^{2}=E_{i,n}^{\ast}( U_{i}^{\ast}U_{j}U_{j}^{\ast}
U_{i}) ^{\otimes n}E_{i,n}+E_{j,n}^{\ast}( U_{j}^{\ast}U_{i}
U_{i}^{\ast}U_{j}) ^{\otimes n}E_{j,n}.
\]
The two hermitian matrices composing $\Delta_{ij,n}^{2}$ are
orthogonal, and
their nonze\-ro eigenvalues are those of $( U_{i}^{\ast
}U_{j}U_{j}^{\ast
}U_{i}) ^{\otimes n}$ and $( U_{j}^{\ast}U_{i}U_{i}^{\ast}
U_{j}) ^{\otimes n}$, respectively. Hence,
%
%e4.8 #&#
\begin{eqnarray}\label{ev-estim-3-a}
\lambda_{\max}( \Delta_{ij,n}^{2}) & = & \max\{
\lambda_{\max}( U_{i}^{\ast}U_{j}U_{j}^{\ast}U_{i}
)^{\otimes
n},\lambda_{\max}(
U_{j}^{\ast}U_{i}U_{i}^{\ast}U_{j})^{\otimes n}\}
\nonumber\\[-8pt]\\[-8pt]
& = & \max\{
\lambda_{\max}^{n}(U_{i}^{\ast}U_{j}U_{j}^{\ast} U_{i})
,\lambda_{\max}^{n}(U_{j}^{\ast}U_{i}U_{i}^{\ast} U_{j})
\}.\nonumber
\end{eqnarray}
Let $P_{i}=U_{i}U_{i}^{\ast}$ be the projection operator onto the space
$\operatorname{supp}(\rho_{i}) =\operatorname{span}(U_{i})$. Note that
$U_{i}^{\ast}P_{j}U_{i}$ and $P_{i}P_{j}P_{i}$ have the same set of nonzero
eigenvalues, hence by Lemma \ref{lem-eigenvalues-itproj} below and Condition
\ref{condLI} we have $\lambda_{\max}( U_{i}^{\ast}P_{j}U_{i})
<1$ and
$\lambda_{\max}( U_{j}^{\ast}P_{i}U_{j}) <1$. It follows
\begin{eqnarray*}
\lambda_{\max}^{n}( U_{i}^{\ast}P_{j}U_{i}) &
\rightarrow&
0 \qquad\mbox{as }n\rightarrow\infty,\\
\lambda_{\max}^{n}( U_{j}^{\ast}P_{i}U_{j}) &
\rightarrow&
0\qquad \mbox{as }n\rightarrow\infty,
\end{eqnarray*}
hence by (\ref{ev-estim-3-a}) $\lambda_{\max}( \Delta
_{ij,n}^{2})
\rightarrow0$. Thus, (\ref{delta-ij-bound-a}) is established.
\end{pf}
%
%le3 #&#
\begin{lemma}
\label{lem-eigenvalues-itproj} Let $\mathcal{L}_{0},\mathcal{L}_{1}$
be linear
subspaces of $\mathbb{C}^{d}$ and $P_{0},P_{1}$ be the corresponding
projection operators. Then
$\mathcal{L}_{0}\cap\mathcal{L}_{1}=\{
0\} $ if and only
if
\[
\lambda_{\max}( P_{0}P_{1}P_{0})
<1.
\]
\end{lemma}
\begin{pf}
It is obvious that always $\lambda_{\max}(
P_{0}P_{1}P_{0})
\leq1$, so it suffices to prove that $\mathcal{L}_{0}\cap\mathcal{L}_{1}
\neq\{ 0\} $ is equivalent to $\lambda_{\max}( P_{0}
P_{1}P_{0}) =1$. Assume there exists\vadjust{\goodbreak} $x\in\mathcal{L}_{0}
\cap\mathcal{L}_{1}$, $\Vert x\Vert>0$, then
$P_{i}x=x$, $i=0,1$
and hence $P_{0}P_{1}P_{0}x=x$ so that $\lambda_{\max}( P_{0}P_{1}
P_{0}) =1$. For the other direction, assume
%
%e4.9 #&#
\begin{equation} \label{ass-ev-1}
\lambda_{\max}( P_{0}P_{1}P_{0}) =1.
\end{equation}
Then there exist $v_{0}\,{\in}\,\mathbb{C}^{d}$, $\Vert
v_{0}\Vert\,{=}\,1$ such that $\langle
v_{0}|P_{0}P_{1}P_{0}|v_{0}\rangle\,{=}\,1$. Here $\Vert
P_{0}v_{0}\Vert\leq1$ by the properties of projections. Assume
$\Vert P_{0}v_{0}\Vert<1$. Then for $u_{0}=P_{0}v_{0}$ we
have
\[
\langle v_{0}|P_{0}P_{1}P_{0}|v_{0}\rangle=\langle
u_{0}|P_{1}|u_{0}\rangle<1,
\]
which contradicts the assumption (\ref{ass-ev-1}). Hence, we must have
$\Vert P_{0}v_{0}\Vert=1$ and hence $v_{0}\in\mathcal
{L}_{0}$ and
$P_{0}v_{0}=v_{0}$. Then
\[
1=\langle v_{0}|P_{0}P_{1}P_{0}|v_{0}\rangle=
\langle
v_{0}|P_{1}|v_{0}\rangle,
\]
which implies $v_{0}\in\mathcal{L}_{1}$ by an analogous reasoning. Hence,
$v_{0}\in\mathcal{L}_{0}\cap\mathcal{L}_{1}$ where $\Vert v_{0}
\Vert=1$, hence $\mathcal{L}_{0}\cap\mathcal{L}_{1}\neq
\{
0\} $.
\end{pf}
\begin{pf*}{Proof of Theorem \ref{thmmultiple-Chernoff}}
We utilize the detector
constructed in Algorithm \ref{alg-GS}, applied to the tensor product case
$\Sigma=\Sigma^{\otimes n}$; call this detector $E^{(n)}$. Lemma \ref
{lemasymp-norm-of-Gram-matrices} implies that the set $\mathcal{V}_{n}$
is a linearly independent set for sufficiently large $n$. As a consequence,
when Lemma \ref{lem-general-algorithm} is applied to the tensor
product set
$\Sigma^{\otimes n}=\{ \rho_{1}^{\otimes n},\ldots,\rho
_{r}^{\otimes n
}\} $, the matrix $\Gamma_{N}$ occurring there equals
$\mathring
{\Gamma}_{n}$ up to a~rearrangement and $\lambda_{\min}(
\Gamma
_{N}) =\lambda_{\min}(\mathring{\Gamma}_{n})$. We find from
(\ref{estav-err}) that
%
%e4.10 #&#
\begin{eqnarray} \label{reasoning-as-in-1}
\operatorname{Err}\bigl(E^{(n)}\bigr) & \leq & \lambda_{\min}^{-1}(\mathring{\Gamma}
_{n}) r^{-1}\sum_{1\leq i,j\leq r,j\not=i}\inf_{s\in[0,1]}
\operatorname{tr}[ ( \rho_{i}^{\otimes n}) ^{1-s}(
\rho_{j}^{\otimes n}) ^{s}]\nonumber\\[-8pt]\\[-8pt]
& = & r^{-1}\bigl( 1+o(1)\bigr) \sum_{1\leq i,j\leq r,j\not=i}\Bigl(
\inf_{s\in[0,1]}\operatorname{tr}[ \rho_{i}^{1-s}\rho_{j}
^{s}] \Bigr)^{n}.\nonumber
\end{eqnarray}
Recall the definition (\ref{defq-Chernoff-dist}) of the pairwise
quantum Chernoff bound $\xi_{\mathrm{QCB}}(\rho_{i},\allowbreak\rho_{j})$; then
%
%e4.11 #&#
\begin{equation}\label{reasoning-as-in-2}
\operatorname{Err}\bigl(E^{(n)}\bigr)\leq r^{-1}\bigl( 1+o(1)\bigr) \sum_{1\leq
i,j\leq
r,j\not=i}\exp( -n\xi_{\mathrm{QCB}}(\rho_{i},\rho_{j})).
\end{equation}
Taking $\log$ of both sides and dividing by $n$, the limit of the
right-hand side above is determined by the smallest of the
$\xi_{\mathrm{QCB}}(\rho_{i},\rho_{j} )$, which according to~(\ref{defq-multiple-Chernoff-dist})
coincides with
$\xi_{\mathrm{QCB}}(\Sigma)$. The theorem follows.
%%%%%%%%%%%%%%%%%%%%%%%%%%%%%%%%%%%%%%%%%%%%%%%%%%%%%%%%%%%%%%%%
\end{pf*}

%s5 #&#
\section{Commuting states}\label{seccommute}

Suppose all the density matrices $\rho_{i}$ are commuting:
$\rho_{i}\rho_{j}=\rho_{j}\rho_{i}$ for all $i,j\in[1,r]$. Then
the $\rho_{i}$ have a common set of eigenvectors $v_{j}$,
$j\in[1,d]$. The spectral decompositions
(\ref{spec-decompose-rho}) now are
\[
\rho_{i}=\sum_{j=1}^{d}\lambda_{i,j}|v_{j}\rangle\langle v_{j}|,\qquad
i\in[1,r].
\]
Also, w.l.o.g., by applying a unitary transformation, we can assume
that all~$\rho_{i}$ are diagonal matrices and $v_{j}$ is a
canonical\vadjust{\goodbreak}
basis vector of $\mathbb{C}^{d}$. Then the set of eigenvalues of
$\rho_{i}$ represents a probability distribution $P_{i}$ on a finite
sample space $\Omega$, $\#\Omega=d$, where each $\omega\in\Omega$ can
be identified with one of the projections $|v_{j}\rangle\langle
v_{j}|$.

With this identification, Algorithm \ref{alg-GS} reduces essentially
to Algorithm~\ref{alg-ML}. Indeed, in the orthogonalization step
(\ref{G-S-orthogonalize-crucial}), the newly appearing unit vector
$v_{i^{\ast}j^{\ast}}$ in step $s$ is one of the basis vectors
$v_{j}$. By induction, it follows that the constructed basis
$e_{1},\ldots,e_{d}$ coincides with $v_{1},\ldots,v_{d}$ up to
possible reindexing and change of sign. Thus, the classical decision
rule $\varphi$ found in Algorithm \ref{alg-GS} on the sample space
elements $|e_{j}\rangle\langle e_{j}|$ is equivalent to a~decision
rule on $\Omega$, constructed according to Algorithm \ref{alg-ML}, and
the latter is a~maximum likelihood rule. The ML rule is not unique in
general; in case of nonuniqueness, any version may result from
Algorithm
\ref{alg-ML}, according to the choice of a maximal entry in step (i).

In Lemma \ref{lem-general-algorithm}, $\Gamma_{L}$ is the Gram matrix
pertaining to $\{ v_{j}\} _{j=1}^{d}$, that is unity.
Thus, we
obtain
\begin{eqnarray*}
\operatorname{Err}(E)&\leq& r^{-1}\sum_{1\leq i,j\leq r,j\not=i}\inf
_{s\in
[0,1]}\operatorname{tr}[\rho_{i}^{1-s}\rho_{j}^{s}]
\\ &=& r^{-1}\sum_{1\leq
i,j\leq r,j\not=i}\inf_{s\in[0,1]}\sum_{\omega\in\Omega}
P_{i}^{1-s}( \omega) P_{j}^{s}( \omega)
\end{eqnarray*}
and reasoning further as in (\ref{reasoning-as-in-1}) and (\ref{reasoning-as-in-2}
), we have thus reproduced the attainability result for the multiple classical
Chernoff bound (cf. (\ref{class-multip-Chern}) and
\cite{Salikh-73,Salikh-97}).

%s6 #&#
\section{A near optimal rate in the general case}\label{secnear-optimal}

We establish that, as stated in Theorem \ref{thmexp-decay}, in the
general case of a finite number of quantum hypotheses there exist
quantum tests that achieve an error exponent equal to the generalized
quantum Chernoff distance up to a factor $1/3$.

To construct the detector attaining the exponential bound in the
general case, we will modify Algorithm \ref{alg-GS} such that it
assumes certain density matrices $\tilde{\rho}_{i}$, which represent
$\varepsilon$-perturbations of embeddings of the original $\rho_{i}$
into a higher-dimensional space $\mathbb{C}^{D}$, $D>d$. These states
$\tilde{\rho}_{i}$ are not observable; the detector will be applied to
the extensions of $\rho_{i}$, which are observable.

Set $D=(r+1)d$ and consider the $k$th canonical unit vector $f_{k}$
in $(r+1)d$-dimensional space $\mathbb{C}^{D}$. Reindex the basis
vectors $f_{k}$ such that $f_{i,j}=f_{(i-1)d+j}$ for $(
i,j) \in[1,r+1]\times
[1,d]$ and define subspaces
\[
S_{i}=\operatorname{span}\{ f_{i,j}\} _{j=1}^{d}.
\]
Then $\mathbb{C}^{D}$ is a direct sum $\mathbb{C}^{D}=
%TCIMACRO{\dbigoplus _{i=1}^{r+1}}
%BeginExpansion
\bigoplus _{i=1}^{r+1}
%EndExpansion
S_{i}$ where all $S_{i}$ are isomorphic to~$\mathbb{C}^{d}$. Let the
operator $F$ represent the canonical embedding
$F\dvtx\mathbb{C}^{d}\rightarrow S_{1}$. Recall the spectral
representation (\ref{spec-decompose-rho}) of $\rho_{i}$ with
eigenvectors $v_{ij}\in\mathbb{C}^{d}$; setting $u_{i,j}=Fv_{ij}$, we
may equivalently assume that instead of $\rho_{i}$ we\vadjust{\goodbreak} measure a~\mbox{$D\times D$} density matrix $\rho_{0,i}$ having spectral representation
\[
\rho_{0,i}=\sum_{i=1}^{d}\lambda_{i,j}\vert u_{i,j}
\rangle
\langle u_{i,j}\vert.
\]
For $\varepsilon\in(0,1)$ and $\delta_{\varepsilon}=(
1-\varepsilon
^{2}) ^{1/2}$, define vectors
\[
\tilde{u}_{i,j}:=\delta_{\varepsilon}u_{i,j}+\varepsilon f_{i+1,j}
\]
for $( i,j) \in J=[1,r]\times[1,d]$. Then, since
$\langle u_{i,j} |f_{i+1,j}\rangle=0$, the vectors
$\tilde{u}_{i,j}$ are unit vectors; define density matrices
%
%e6.1 #&#
\begin{equation} \label{rho-i-tilde-def}
\tilde{\rho}_{i}=\sum_{i=1}^{d}\lambda_{i,j}\vert\tilde{u}
_{i,j}\rangle\langle\tilde{u}_{i,j}\vert,\qquad
i\in[ 1,r].
\end{equation}
Relative to this set of density matrices on $\mathbb{C}^{D}$, satisfying
%
%e6.2 #&#
\begin{equation}\label{approx-trace-invariance}
\operatorname{tr}[\tilde\rho_i^{1-s}\tilde\rho_j^{s}]=
\delta_{\varepsilon}^4 \operatorname{tr}[\rho_i^{1-s}\rho_j^s]
\end{equation}
construct a detector according to (\ref{pvm-construct}) and Algorithm
\ref{alg-GS}, and call this $\tilde{E}_{\varepsilon}$. Then each
$\tilde{E}_{\varepsilon,i}$ is a projection matrix in $\mathbb{C}^{D}$
and $\sum_{i=1}^{r}\tilde
{E}_{\varepsilon,i}={\mathbf{1}}_{D}$. Define now~$E_{\varepsilon,i}$
as the upper $d\times d$ submatrix of
$\tilde{E}_{\varepsilon,i}$. Then $E_{\varepsilon,i}$ is a positive
matrix and $\sum_{i=1}^{r}E_{\varepsilon,i}={\mathbf{1}}_{d}$, so
that
%
%e6.3 #&#
\begin{equation}\label{E-epsilon-tilde-def}
E_{\varepsilon}:=\{ E_{\varepsilon,i}\}_{i=1}^{r}
\end{equation}
constitutes a POVM in $\mathbb{C}^{d}$.

It should be noted that $E_{\varepsilon,i}$ are not projections, that is,
$E_{\varepsilon}$ is a general POVM but not a PVM, contrary to the
detector constructed in Algorithm \ref{alg-GS}. However,
$E_{\varepsilon}$ results from a PVM $\tilde{E}_{\varepsilon}$ in a
higher-dimensional space by taking submatrices. This relationship
holds between POVMs and PVMs in general, on the basis of Naimark's
theorem; cf. Parthasarathy \cite{parth-extr} for a discussion.
%
%le4 #&#
\begin{lemma}\label{lemepsilon}
Let $\Sigma=\{\rho_{i}\}_{i=1}^{r}$ be an arbitrary set of density
matrices on~$\mathbb{C}^{d}$. For sufficiently small $\varepsilon>0$, the
detector $E_{\varepsilon}$
constructed in (\ref{E-epsilon-tilde-def}) fulfills
%
%e6.4 #&#
\begin{equation} \label{est-av-err-E-epsilon-tilde}
\operatorname{Err}(E_{\varepsilon})\leq r^{-1}\biggl( 2\varepsilon
+\varepsilon
^{-2}\sum_{1\leq i,j\leq r,j\not
=i}\inf_{s\in[0,1]}\operatorname{tr}
[\rho_{i}^{1-s}\rho_{j}^{s}]\biggr).
\end{equation}
\end{lemma}
\begin{pf}
$\!\!\!$Consider the Gram matrix $\tilde{\Gamma}_{J}$ of the set of vectors
\mbox{$\{\tilde{u}_{i,j},( i,j) \in J\} $}. Since for $(
i,j) \in J$ and $( k,l) \in J$ we have
\[
\langle\tilde{u}_{i,j}|\tilde{u}_{k,l}\rangle=\delta
_{\varepsilon}^{2}\langle u_{i,j}|u_{k,l}\rangle
+\varepsilon^{2}\langle f_{i+1,j}|f_{k+1,l}\rangle
\]
it follows that $\tilde{\Gamma}_{J}$ is a convex combination of two
Gram matrices, which implies that
\[
\lambda_{\min}( \tilde{\Gamma}_{J}) \geq\varepsilon^{2}.
\]
Hence, $\{ \tilde{u}_{i,j},( i,j) \in J\} $ is a set of $rd$ linearly
independent vectors in $\mathbb{C}^{D}$. Since Algorithm \ref{alg-GS}
eliminates from $V_{1}(\Sigma)$ all eigenvectors pertaining to zero
eigenvalues, the sequence $V_{1}(\Sigma)$ of length $L$ contains
exactly the vectors $\{ \tilde{u}_{i,j},( i,j) \in J\} $ pertaining to
nonzero $\lambda_{i,j}$ in (\ref{rho-i-tilde-def}). Their\vspace*{1pt}
full Gram matrix~$\Gamma_{L}$ as given by (\ref{Gamma-s-def}) for $s=L$
is a submatrix of $\tilde{\Gamma}_{J}$ (after rearrangement) and hence
also fulfills
%
%e6.5 #&#
\begin{equation}\label{min-eigenval-epsilon-algorithm}
\lambda_{\min}( \Gamma_{L}) \geq\varepsilon^{2}.
\end{equation}
Consider the error probability of the POVM $E_{\varepsilon}$
%
%e6.6 #&#
\begin{eqnarray} \label{right-of}
\operatorname{Err}(E_{\varepsilon}) & = & 1-r^{-1}
\sum _{i=1}^{r}
\operatorname{tr}[ \tilde{E}_{\varepsilon,i}\rho_{0,i}]
\nonumber\\[-8pt]\\[-8pt]
& = & 1-r^{-1}
\sum _{i=1}^{r}
\operatorname{tr}[ \tilde{E}_{\varepsilon,i}\tilde{\rho}_{i}]
+r^{-1}
\sum _{i=1}^{r}
\operatorname{tr}[ \tilde{E}_{\varepsilon,i}( \tilde{\rho}
_{i}-\rho_{0,i}) ] .\nonumber
\end{eqnarray}
Now according to Lemma \ref{lem-general-algorithm},
(\ref{min-eigenval-epsilon-algorithm}), and
(\ref{approx-trace-invariance}) we have
\begin{eqnarray*}
1-r^{-1}
\sum _{i=1}^{r}
\operatorname{tr}[ \tilde{E}_{\varepsilon,i}\tilde{\rho}_{i}]
& \leq & \varepsilon^{-2} r^{-1}\sum_{1\leq i<j\leq r}\inf
_{s\in[0,1]}\operatorname{tr}[\tilde{\rho}_{i}^{1-s}\tilde{\rho}
_{j}^{s}]\\
& \leq &
\varepsilon^{-2} r^{-1} \sum_{1\leq i<j\leq r}\inf_{s\in
[0,1]}\operatorname{tr}[\rho_{i}^{1-s}\rho_{j}^{s}].
\end{eqnarray*}
For the second
term on the right-hand side of (\ref{right-of}) note that
\[
\tilde{\rho}_{i}-\rho_{0,i}=\sum_{j=1}^{d}\lambda_{i,j}(
\vert
\tilde{u}_{i,j}\rangle\langle\tilde{u}_{i,j}
\vert
-\vert u_{i,j}\rangle\langle u_{i,j}\vert
).
\]
Here we have
\begin{eqnarray*}
&& \vert\tilde{u}_{i,j}\rangle\langle\tilde{u}
_{i,j}\vert-\vert u_{i,j}\rangle\langle u_{i,j}
\vert\\
&&\qquad =\vert\delta_{\varepsilon}u_{i,j}+\varepsilon
f_{i+1,j}\rangle
\langle\delta_{\varepsilon}u_{i,j}+\varepsilon f_{i+1,j}
\vert
-\vert u_{i,j}\rangle\langle u_{i,j}\vert\\
&&\qquad = -\varepsilon^{2}\vert u_{i,j}\rangle\langle
u_{i,j}\vert\\
&&\qquad\quad{}+\delta_{\varepsilon}\varepsilon\vert u_{i,j}
\rangle\langle f_{i+1,j}\vert+\delta
_{\varepsilon
}\varepsilon\vert f_{i+1,j}\rangle\langle
u_{i,j}\vert
+\varepsilon^{2}\vert f_{i+1,j}\rangle\langle f_{i+1,j}
\vert\\
&&\qquad =\delta_{\varepsilon}\varepsilon\vert
u_{i,j}+f_{i+1,j}\rangle
\langle u_{i,j}+f_{i+1,j}\vert\\
&&\qquad\quad{} - ( \delta
_{\varepsilon
}\varepsilon-\varepsilon^{2}) ( \vert
u_{i,j}\rangle
\langle u_{i,j}\vert+\vert f_{i+1,j}\rangle
\langle f_{i+1,j}\vert)\\
&&\qquad\quad{}  - 2 \varepsilon^2 \vert u_{i,j}\rangle
\langle u_{i,j}\vert.
\end{eqnarray*}
Since the matrix
\[
( \delta_{\varepsilon}\varepsilon-\varepsilon^{2})
(
\vert u_{i,j}\rangle\langle u_{i,j}\vert
+\vert
f_{i+1,j}\rangle\langle f_{i+1,j}\vert) +
2\varepsilon^2 \vert u_{i,j}\rangle
\langle u_{i,j}\vert
\]
is positive for sufficiently small $\varepsilon$, we have
\[
\vert\tilde{u}_{i,j}\rangle\langle\tilde{u}_{i,j}
\vert-\vert u_{i,j}\rangle\langle
u_{i,j}\vert
\leq\delta_{\varepsilon}\varepsilon\vert
u_{i,j}+f_{i+1,j}\rangle
\langle u_{i,j}+f_{i+1,j}\vert
\]
consequently
\begin{eqnarray*}
\operatorname{tr}[ \tilde{E}_{\varepsilon,i}( \tilde{\rho}
_{i}-\rho_{0,i}) ] & \leq & \sum_{j=1}^{d}\lambda_{i,j}
\operatorname{tr}[ \tilde{E}_{\varepsilon,i}( \delta
_{\varepsilon}\varepsilon\vert u_{i,j}+f_{i+1,j}\rangle
\langle u_{i,j}+f_{i+1,j}\vert) ] \\
& \leq & \sum_{j=1}^{d}\lambda_{i,j}\operatorname{tr}[ (
\delta_{\varepsilon}\varepsilon\vert u_{i,j}+f_{i+1,j}
\rangle
\langle u_{i,j}+f_{i+1,j}\vert) ] \\
& = &\delta_{\varepsilon}\varepsilon\sum_{j=1}^{d}\lambda_{i,j}\cdot
2\leq2\varepsilon.
\end{eqnarray*}
\upqed\end{pf}
\begin{pf*}{Proof of Theorem \ref{thmexp-decay}}
We denote
the factor of $\varepsilon^{-2}$ in (\ref{est-av-err-E-epsilon-tilde})
by $K_1$, and in the $n$-fold tensor product case, where $\rho_{i}$ is
replaced by $\rho_{i}^{\otimes n}$, by $K_n$, respectively. To find
the best upper bound in
(\ref{est-av-err-E-epsilon-tilde}), we minimize the expression
$2\varepsilon+\varepsilon^{-2}K_n$ in~$\varepsilon$. The solution is
$\varepsilon=K_n^{1/3}$ and
the value at the minimum is $3K_n^{1/3}$. Since $K_n$ tends
to zero as $n$ goes to infinity it is ensured that for sufficently
large $n$, the value $K_n^{1/3}$ is small enough to satisfy the
condition of
Lemma \ref{lemepsilon}. Thus from
(\ref{est-av-err-E-epsilon-tilde}) we obtain
\[
\operatorname{Err}\bigl(E_{\varepsilon}^{(n)}\bigr)\leq 3r^{-1}\biggl( \sum_{1\leq
i,j\leq
r,j\not=i}\inf_{s\in[0,1]}\operatorname{tr}[(\rho
_{i}^{\otimes n})^{1-s}(\rho_{j}^{\otimes n}
)^{s}]\biggr) ^{1/3},
\]
where $E^{(n)}_{\varepsilon}$ denotes the respective detectors in the
tensor product case $\Sigma^{\otimes n}$. It follows
\begin{eqnarray*}
\frac{1}{n}\log\operatorname{Err}\bigl(E^{(n)}\bigr) & \leq & \frac{1}{3}\frac
{1}{n}\log\biggl(
\sum_{1\leq i,j\leq r,j\not=i}\inf_{s\in[0,1]}\operatorname{tr}
[ ( \rho_{i}^{\otimes n})^{1-s}(
\rho_{j}^{\otimes n})^{s}] \biggr) +o(1)\\
& = & \frac{1}{3}\log\xi_{\mathrm{QCB}}( \Sigma) +o(1),
\end{eqnarray*}
which proves our claim.
\end{pf*}

%suskaldyti doi

% imsref loaded by lrinkeviciute, 2011-12-22 15:24:39
% imsref loaded by lrinkeviciute, 2011-12-22 15:38:25

\printaddresses

\end{document}